\documentclass[pra,twocolumn,preprintnumbers,amsmath,amssymb,showpacs,nofootinbib]{revtex4}
\usepackage{graphicx}
\usepackage{graphics}
\usepackage{psfrag}
\usepackage{hyperref}
 \usepackage{color}

\newcommand{\beq}{\begin{equation}}
\newcommand{\eeq}{\end{equation}}

\begin{document}

\title{Renormalization group study of the four-body problem}
\author{Richard Schmidt}\email{richard.schmidt@ph.tum.de}
\author{Sergej Moroz}\email{s.moroz@thphys.uni-heidelberg.de}
\affiliation{$^*$Physik Department, Technische Universit\"at M\"unchen, James-Franck-Strasse, D-85748 Garching, Germany\\$^\dagger$Institut f\"{u}r Theoretische Physik, Universit\"at Heidelberg, Philosophenweg 16, D-69120 Heidelberg, Germany}

\begin{abstract}
We perform a renormalization group analysis of the non-relativistic four-boson problem by means of a simple model with pointlike three- and four-body interactions. We investigate in particular the region where the scattering length is infinite and all energies are close to the atom threshold. We find that the four-body problem behaves truly universally, independent of any four-body parameter. Our findings confirm the recent conjectures of Platter et al. and von Stecher et al. \cite{Platter04,Hammer07,Stecher09} that the four-body problem is universal, now also from a renormalization group perspective. We calculate the corresponding relations between the four- and three-body bound states, as well as the full bound state spectrum and comment on the influence of effective range corrections.
\end{abstract}

\pacs{03.65.Nk, 34.50.-s,11.10.Hi}

\maketitle

\section{Introduction} \label{introduction}
During the last decade few-body physics experienced a renewed interest due to the advent of experiments with ultracold atomic gases. Whereas the study of few-body physics in nuclear systems is hindered by the large complexity of the interparticle potentials, the interactions in ultracold atomic gases are describable to high accuracy with very simple short-range models. In addition, ultracold atomic gases become even more attractive as an ideal theoretical and experimental playground since they do not only offer excellent experimental control but also the amazing possibility of tuning the two-body interaction strength over a wide range using so-called Feshbach resonances \cite{Chin08}.\\
This made it possible, that, about forty years after V. Efimov's seminal prediction \cite{Efimov} of the existence of universal three-body bound states in systems with large two-body interactions, first evidence in favor of the presence of these states had been found in the remarkable experiment by Kraemer et al. in 2006 \cite{Kraemer06}. In his work, Efimov predicted the existence of infinitely many trimer states for infinitely large scattering length where the two-body interaction is just on the verge of having a bound state. The energy levels form a geometric spectrum and the three-body system is found to be universal in the sense that apart from the s-wave scattering length $a$ only one piece of information about the three-body system enters in the form of a so-called three-body parameter \cite{BH}.
The findings of Kraemer et al. stimulated extensive activity in the field of three-body physics, both experimentally  \cite{Knoop09,Ottenstein08,Huckans09,Zaccanti09,Barontini09} and theoretically; for recent reviews on also the latter see \cite{BH, Platter09} and references therein. As a result, the Efimov effect in three-body systems is a well-understood phenomenon today.\\
The next natural step is to raise the question what the physics of four interacting particles may be. Early attempts towards an understanding of this system were made in the context of nuclear physics using a variety of approaches \cite{Yakubovsky67,Tjon75, Gibson76,Fonseca76}. Also the four-body physics of $^4\textrm{He}$ atoms has been investigated in much detail, for an overview see, e.~g. \cite{HeRevs}. The simpler four-body physics of fermions with two spin states, relevant for the dimer-dimer repulsion, has also been studied \cite{Petrov03}.\\
In their pioneering work, Platter and Hammer, et al. \cite{Platter04,Hammer07} investigated the four-body problem using effective interaction potentials and made the conjecture that the four-boson system exhibits universal behavior. They also found that no four-body parameter is needed for a self-consistent renormalization of the theory. Calculating the energy spectrum of the lowest bound states in dependence on the scattering length $a$ the existence of two tetramer (four-body bound) states associated with each trimer was conjectured.\\
Recently, von Stecher, D'Incao, and Greene  \cite{Stecher09, DIncao09} investigated the four-body problem in a remarkable quantum mechanical calculation. They found that the Efimov trimer and tetramer states always appear as sets of states with two tetramers associated with each of the trimer levels and calculated the bound state energy spectrum of the lowest few sets of states. The calculation suggests that the energy levels within one set of states are related to each other by universal ratios, which were obtained from the behavior of these lowest sets of states. In accordance with the results of Platter et  al. \cite{Platter04,Hammer07} the absence of any four-body parameter was also demonstrated. In order to find experimental evidence of the tetramer states extremely precise measurements are required. Remarkably, Ferlaino et al. were able to observe signatures of the lowest two of the tetramer states in a recent experiment \cite{Ferlaino09}.\\
While the calculations by Platter et al. \cite{Platter04,Hammer07} and von Stecher et al. \cite{Stecher09,DIncao09} rely on quantum mechanical approaches, in this work we want to shed light onto the four-body problem from a different perspective. A lot of insight into the three-body problem had been gained from effective field theory and renormalization group (RG) methods \cite{BH,Bedaque,MFSW,Platter09} and it is desirable to apply these also to the four-body problem. In this paper we will make a first step towards such a description that is complementary to the previous quantum mechanical approaches.\\
\begin{figure}[t]
\begin{center}
\includegraphics[width=\columnwidth]{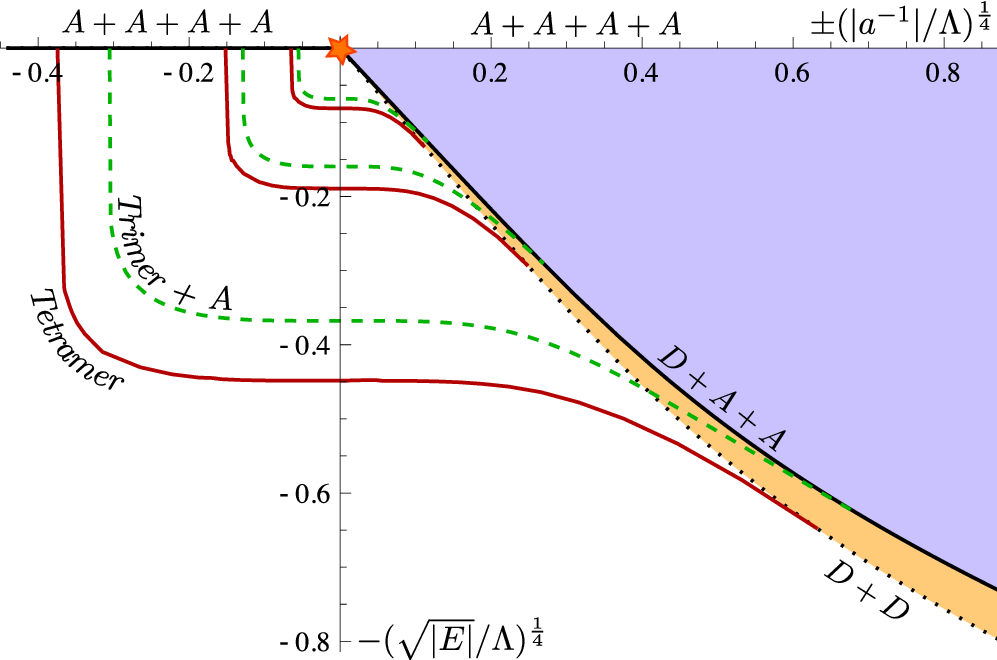}
\end{center}
\vskip -0.7cm \caption{(Color online) The generalized Efimov plot for four identical bosons. Here, we plot the energy levels of the various bound states as a function of the inverse s-wave scattering length $a$ as numerically calculated in our approximative, effective theory. In order to improve the visibility of the energy levels we rescale both the dimensionless energy $E/\Lambda^2$ and the dimensionless inverse scattering length $a^{-1}/\Lambda$ where $\Lambda$ denotes the UV cutoff of our model. Also, we only show the first three sets of Efimov levels. The solid black line denotes the atom-atom-dimer threshold, while the dotted black line gives the dimer-dimer threshold. In the three-body sector one finds the well known spectrum of infinitely many Efimov trimer states (green, dashed) which accumulate at the unitarity point $E_\psi=a^{-1}=0$, indicated by the orange star. In our pointlike approximation the four-body sector features a single tetramer (solid, red) associated with each trimer state.}
\label{4BSpectrum}
\end{figure}
Of special interest is the further investigation of universality in the four-body system. In this context the so-called unitarity point, illustrated by the star in Fig. \ref{4BSpectrum}, is of particular importance. In this limit not only the scattering length $a$ is infinite but also all binding energies in the problem accumulate at the atom threshold at zero energy. Only at the unitarity point physics becomes truly universal in the sense that for example the ratio between the binding energies of consecutive trimer levels assumes exactly its universal value, $E_{n+1}/E_{n}=\exp({-2\pi/s_0})$, with $s_0\approx 1.00624$ the so-called Efimov parameter. The unitarity point is therefore the most interesting one from a theory point of view. Unfortunately, in the previous calculations only few lowest lying states were determined. The major advantage of the present RG approach is that it allows to investigate analytically the complete spectrum and to address directly the unitarity point in order to extract the universal relations between the three- and four-body bound states in this limit.\\
We investigate the renormalization group behavior of the relevant four-body interactions with an approximate, but simple and physically intuitive model which allows only for pointlike (i.e. momentum independent\footnote{Throughout the paper we will use the term momentum independence to refer to combined spatial momentum  and frequency independence.}) three- and four-body interactions. In the three-body problem universality manifests itself in an RG limit cycle of the three-body coupling. We find that this three-body limit cycle leads in turn to a ``self-sustained'' limit cycle of the four-body sector leaving no room for any four-body parameter.\\
The RG method allows furthermore for computations away from the unitarity point. We calculate the bound state energy spectrum in the pointlike approximation (see Fig. \ref{4BSpectrum}) and investigate how the relations between tetramer and trimer states approach the universal limit as one comes closer to the unitarity point.\\
The paper is structured as follows. In Sec. \ref{method} we introduce the functional renormalization group (FRG) method and set up the microscopic model. Secs. \ref{twobody} and \ref{STBS} are devoted to the FRG analysis of the two- and three-body sector. In Sec. \ref{4B} we discuss the four-body sector and present our numerical results. Our findings are summarized in Sec. \ref{Conclusion}.

\section{Method and definition of the model}\label{method}

In this work we are interested in the computation of the few-body properties, such as the bound state spectrum, of four identical bosons. In a quantum field theory approach the information about these properties can be extracted from the effective action $\Gamma$ which is the generating functional of one-particle irreducible vertex functions $\Gamma^{(n)}$ and which contains all information about a given system. The computation of $\Gamma$ is a very complicated task as quantum (and, as in the case of nonzero density, statistical) fluctuations have to be integrated out on all length and therefore momentum scales $q$. In order to cope with this task we rely on the functional renormalization group \cite{Wetterich93}, for detailed reviews we refer to \cite{Reviews,Pawlowski}.\\
The central quantity of the FRG is a scale dependent effective action functional, the so-called effective flowing action $\Gamma_k$. The effective flowing action $\Gamma_k$, which includes all fluctuations with momenta $q\gtrsim k$, interpolates between the classical action $S$ at some ultraviolet (UV) cutoff scale $k=\Lambda$ and the full quantum effective action $\Gamma$ in the limit $k\to 0$. The underlying idea is similar to Wilson's idea of momentum shell-wise integration of fluctuations. The evolution of $\Gamma_k$ is governed by the Wetterich equation \cite{Wetterich93}, which is an exact, non-perturbative RG equation. It reads
\beq \label{m1}
\partial_k \Gamma_k=\frac{1}{2}\text{Tr}\, \left(\Gamma_k^{(2)}+R_k\right)^{-1}\partial_k R_k,
\eeq
where $\Gamma_k^{(2)}$ is the flowing, full inverse propagator and the trace $\textrm{Tr}$ sums over momentum $\vec q$ and Matsubara frequency $q_0$ as well as the internal degrees of freedom such as species of fields. The dependence on the RG scale $k$ is introduced by the regulator $R_k$. At the UV scale $k=\Lambda$ the effective flowing action $\Gamma_k$ equals the classical action $S$ and as we want to consider dilute atomic gases the UV scale $\Lambda$ is set to be of the order of the inverse Bohr radius $a_0^{-1}$. For most problems, quantum and statistical fluctuations will generate infinitely many terms in $\Gamma_k$. Due to this fact it is in practice impossible to solve Eq. \eqref{m1} exactly. Therefore one has to decide for a truncation of $\Gamma_k$, which in turn corresponds to solving the theory only approximately.\\
In this work we investigate the four-boson problem by approximately solving Eq. (\ref{m1}).
Our truncation for the Euclidean flowing action is given by a simple two-channel model
\begin{eqnarray}\label{m2}
\Gamma_k&=&\int_{x}\{\psi^{*}(\partial_{\tau}-\Delta+E_\psi)\psi\notag\\
&+&\phi^{*}\left(A_{\phi}(\partial_{\tau}-\frac{\Delta}{2})+m_{\phi}^{2} \right)\phi +\frac{h}{2}(\phi^{*}\psi\psi+\phi \psi^{*}\psi^{*})\notag\\
&+&\lambda_{AD}\phi^{*}\psi^{*}\phi\psi +\lambda_{\phi}(\phi^{*}\phi)^{2}\notag\\
&+&\beta(\phi^{*}\phi^{*}\phi\psi\psi+\phi\phi\phi^{*}\psi^{*}\psi^{*}) +\gamma \phi^{*}\psi^{*}\psi^{*}\phi\psi\psi \},
\end{eqnarray}
where $\Delta$ denotes the Laplace operator and we use the natural, non-relativistic convention $2M=\hbar=1$ with the atom mass $M$. $\psi$ denotes the field of the elementary bosonic atom, while the dimer, the bosonic bound state consisting of two elementary atoms, is represented by the field $\phi\sim\psi\psi$. Both the atom and the dimer field are supplemented with non-relativistic propagators with energy gaps $E_\psi$ and $m_\phi^2$, respectively. In our approximation the fundamental four-boson interaction $\sim\lambda_\psi (\psi^*\psi)^2$ is mediated by a dimer exchange, which yields $\lambda_\psi=-h^2/m_\phi^2$ in the limit of pointlike two-body interactions. The dynamical dimer field $\phi$ allows us to capture essential details of the momentum dependence of the two-body interaction.  We introduce a wave function renormalization factor $A_{\phi}$ for the dimer field in order to take into account an anomalous dimension of the dimer field $\phi$. The only nonzero interaction, present at the microscopic UV scale $k=\Lambda$, is taken to be the Yukawa-type term with the coupling $h$. Together with the microscopic value of $A_\phi$ the Yukawa interaction $h$ at the UV scale can be connected to the effective range $r_\textrm{eff}$ in an effective range expansion. The atom-dimer interaction $\lambda_{AD}$ as well as the various four-body interactions $\lambda_\phi$, $\beta$, and $\gamma$ vanish at the UV scale and are built up via quantum fluctuations during the RG flow.\\
At this stage we want to emphasize the meaning of the term pointlike approximation which must not be confused with the notion of a zero-range (contact) model. Consider for example the two-body contact interaction $\sim\lambda_\psi (\psi^*\psi)^2$, which has no momentum dependence on the UV scale, $k=\Lambda$.  In order to describe the scattering of two particles in quantum mechanics one proceeds by solving the two-body Schr\"odinger equation. From this one obtains the well-known result for the zero-range s-wave scattering amplitude $f_0(p)=(-a^{-1}-i p)^{-1}$ (with $p=|\vec p |$ denoting the momentum of the colliding particles). The scattering amplitude becomes momentum dependent. In the RG approach one deals with the effective vertex $\lambda_\psi$ which varies with the RG scale $k$. On the UV ($k=\Lambda$) scale $\lambda_\psi$ is momentum independent. When including more and more quantum fluctuations -- meaning lowering the RG scale $k$ from $\Lambda$ to eventually $k=0$ -- the effective vertex function $\lambda_\psi$ assumes a momentum dependence which in the IR limit $k=0$ is equivalent to the result for $f_0$ in the zero-range model. In a pointlike approximation one ignores this generated momentum dependence. In the simple model Eq. \eqref{m2} the three- and four-body sector is treated strictly in the pointlike approximation. However, in the two-body sector the momentum dependence of effective vertex $\lambda_\psi$ is captured by the exchange of the dynamic (i.e. momentum dependent) dimer propagator, such that the two-body sector is treated beyond the pointlike approximation.\\
In the general case of nonzero density and temperature one works in the Matsubara formalism and the integral in Eq. (\ref{m2}) sums over homogenous three-dimensional space and over imaginary time $\int_{x}=\int d^{3}x \int_{0}^{1/T}d\tau$. Although our method allows us to tackle a full, many-body problem at finite temperature in this way, we are interested solely in the few-body (vacuum) physics in this paper, for which density $n$ and temperature $T$ vanish. For $T=0$, $\int_{x}$ reduces to an integral over infinite space and time. Our truncation (\ref{m2}) is based on the simple structure of the non-relativistic vacuum and, as demonstrated in \cite{DKS, MFSW}, numerous simplifications occur when solving Eq. (\ref{m1}) compared with the general, many-body case. The flowing action (\ref{m2}) has a global $U(1)$ symmetry which corresponds to particle number conservation. In the vacuum limit it is also invariant under spacetime Galilei transformations which restricts the form of the non-relativistic propagators to be functions of $\partial_\tau-\Delta$ for the atoms and $\partial_\tau-\Delta/2$ for the dimers. All couplings present in Eq. \eqref{m2} are allowed to flow during the RG evolution and are taken to be momentum-independent in Fourier space as explained above.\\
Besides the ansatz of $\Gamma_k$ we must choose a suitable regulator function $R_{k}$ in order to solve Eq. (\ref{m1}). Based on our recent treatment of the closely related three-fermion problem \cite{FSMW,FSW}, we choose optimized regulators, 
\begin{eqnarray} \label{m3}
R_{\psi}&=&(k^2-q^2)\theta(k^2-q^2),\notag\\
R_{\phi}&=&\frac{A_{\phi}}{2}(k^2-q^2)\theta(k^2-q^2),
\end{eqnarray}
with $q=|\vec q|$. These regulators are optimized in the sense of \cite{Litim, Pawlowski} and allow to obtain analytical results.

\section{Two-body sector}\label{twobody}
A remarkable and very useful feature of the vacuum flow equations is comprised by a special hierarchy: the flow equations of the $N$-body sector do not influence the renormalization group flows of the lower $N-1$-body sector \cite{MFSW}. For this reason the different $N$-body sectors can be solved subsequently. In this spirit we first solve the two-body sector, then investigate the three-body sector in order to finally approach the four-body problem.

The solution for the two-body sector can be found analytically in our approximation\footnote{Remarkably, the atom inverse propagator (one-body sector) is not renormalized in the non-relativistic vacuum.} (for the analogous problem considering fermions, see \cite{Diehl, FSMW}). The only running couplings in the two-body sector are the dimer gap $m_\phi^2$ and its wave function renormalization $A_\phi$. The flow equations of the two-body sector are shown in terms of Feynman diagrams in Fig. \ref{FigFlows}(a) and read
\begin{eqnarray}\label{m3a}
\partial_t m_\phi^2&=&\frac{h^2}{12\pi^2}\frac{k^5}{(k^2+E_\psi)^2},\notag\\
\partial_t A_\phi&=&-\frac{h^2}{12\pi^2}\frac{k^5}{(k^2+E_\psi)^3},
\end{eqnarray}
where $t=\ln\frac{k}{\Lambda}$. As there are no possible nonzero flow diagrams for the Yukawa coupling $h$, it does not flow in the vacuum limit.\\
The infrared (IR) values of the couplings $h$ and $m_{\phi}^{2}$ can be related to the low-energy s-wave scattering length $a$ via
\beq \label{m4}
a=-\frac{h^2(k=0)}{16\pi m_{\phi}^{2}(k=0,E_\psi=0)}.
\eeq

Knowing the analytical solution of the two-body sector, this relation can be used to fix the initial values of our model. For the UV value of the dimer gap $m_\phi^2$ we find
\beq \label{m4a}
m_{\phi}^{2}(\Lambda)=-\frac{h^2}{16\pi}a^{-1}+\frac{h^2}{12 \pi^2}\Lambda+2E_\psi.
\eeq
The first term fixes the s-wave scattering length according to Eq. \eqref{m4}, while the second term represents a counterterm taking care of the UV renormalization of the two-body sector. Finally, the last term accounts for the fact that the dimer consists of two elementary atoms. Additionally, we choose $A_{\phi}(\Lambda)=1$ which corresponds to the effective range $r_\textrm{eff}=-\frac{64\pi}{h^2}$.\\

The action \eqref{m2} can also be used for a quite accurate description of Feshbach resonances. In this context Eq. \eqref{m2} is referred to as resonance model. In such a model $m_\phi^2$ is proportional to the detuning energy of the molecule in the closed channel with respect to the atom-atom threshold \cite{Kokkelmans} and the coupling $h$ is proportional to the width of the associated Feshbach resonance being a function of the strength of the coupling to the closed channel. The choice  $A_{\phi}(\Lambda)=1$ then corresponds to the so-called characteristic length $r^*=-\frac{1}{2}r_\textrm{eff}$ often used in literature \cite{BH, Zwerger}.\\
 In the limit of large, positive scattering length there exists a universal, weakly bound dimer state. In order to find its binding energy we 
 calculate the pole of the dimer propagator, corresponding to the condition $m_\phi^2(E_\psi,k=0)=0$, which yields in the limit $E_\psi/\Lambda^2\ll 1$
\begin{eqnarray}\label{dimbinding}
E_D=-2 E_\psi&=&-2\left(\frac{h^2}{64\pi}-\sqrt{\frac{h^4}{(64\pi)^2}+\frac{h^2 a^{-1}}{32\pi}}\right)^2\notag\\
&=&-\frac{2}{r_{\textrm{eff}}^2}\left( 1-\sqrt{1-\frac{2 r_\textrm{eff}}{a}}\right)^2.
\end{eqnarray}
In the limit $h\to\infty$, corresponding to $r_\textrm{eff}\to 0$ one recovers the well-known result $E_D=-2/a^2$. The dimer bound state energy is shown as a function of the inverse scattering length in Fig. \ref{4BSpectrum} (black solid line). The deviation from the universal $1/a^{2}$ scaling for large inverse scattering lengths is due to the finite size of $h$ which is
taken to be $h^2/\Lambda=10$ in Fig. \ref{4BSpectrum}. In the regime of small scattering length $a$ one finds a crossover of the behavior of the dimer binding energy which then has the limiting behavior $E_D=4/(a r_\textrm{eff})$.

\section{Three-body sector}\label{STBS}
The bound state spectrum of the three-body sector is much richer than the one of the two-body system. In his seminal papers \cite{Efimov} Efimov showed the existence of an infinite series of three-body bound states for strong two-body interactions. These energy levels exhibit a universal geometric scaling law as one approaches the unitarity point $E_\psi=a^{-1}=0$. Remarkably, these three-body bound trimer states exist even for negative scattering lengths $a$ where no two-body bound state is present; they become degenerate with the three-atom threshold for negative scattering length and merge into the atom-dimer  threshold for positive $a$. An additional three-body parameter is needed in order to determine the actual positions of the degeneracies \cite{Bedaque}. In this section we want to shortly review how Efimov physics can be treated within our approach. For a more detailed account on that matter and an application to the three-component $^6\textrm{Li}$ Fermi gas we refer to \cite{FSMW,MFSW,FSW}.\\

\begin{figure}[t!]
\begin{center}
\includegraphics[width=\columnwidth]{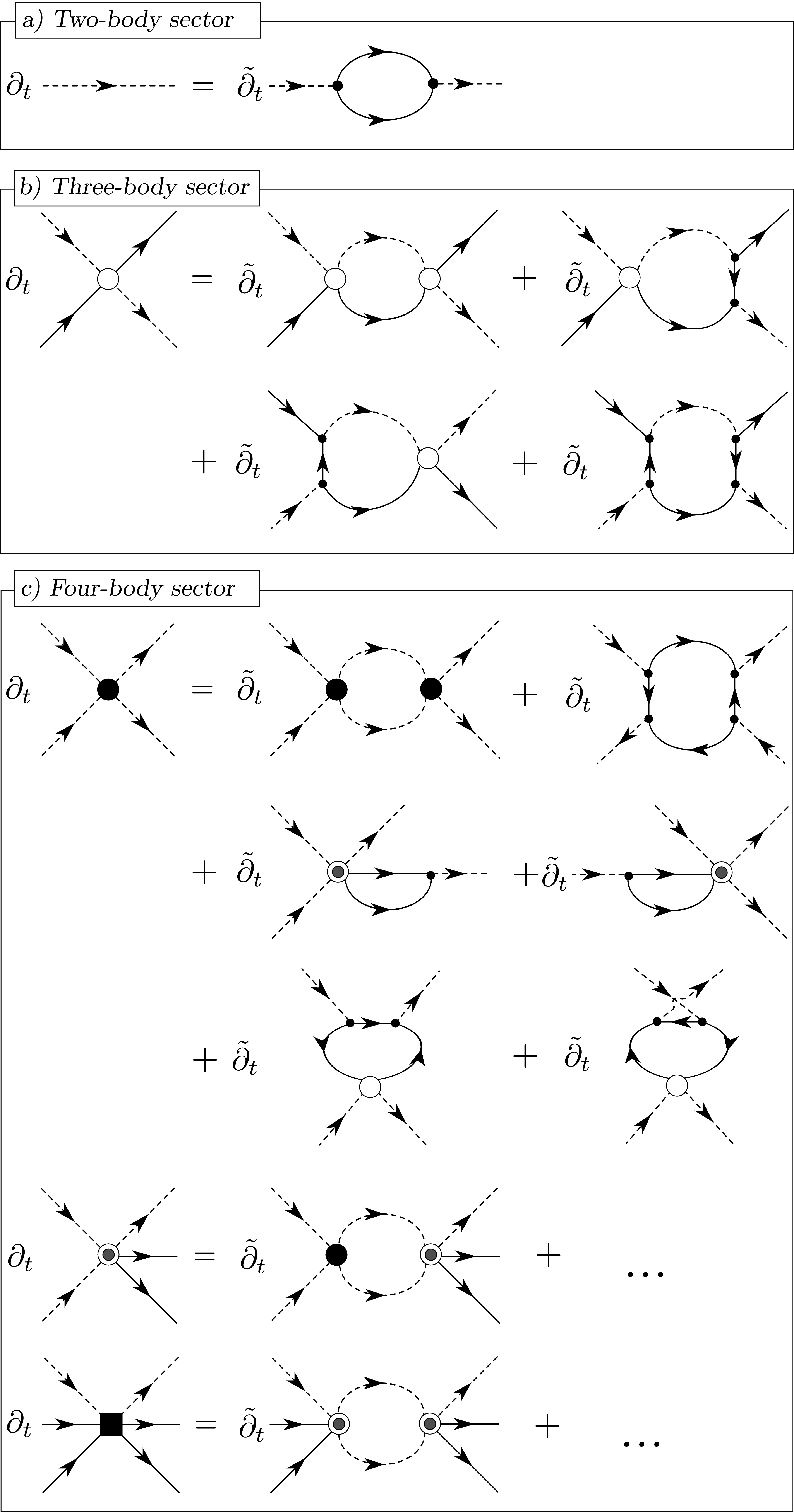}
\end{center}
\caption{The flow equations in terms of Feynman diagrams for the (a) two-body, (b) three-body, and (c) four-body sector. All internal lines denote full, regularized propagators. The scale derivative $\tilde\partial_t$ on the right hand side of the flow equations acts only on the regulators. Solid lines represent elementary bosons $\psi$, while dashed lines denote composite dimers $\phi$. The vertices are: Yukawa coupling $h$ (small black dot), atom-dimer vertex $\lambda_{AD}$ (open circle), dimer-dimer coupling $\lambda_\phi$ (black circle), coupling $\beta$ (two circles), and the atom-atom-dimer vertex $\gamma$ (black square). Due to the large number of diagrams for the latter two vertex functions, we only show two exemplary diagrams. }
\label{FigFlows}
\end{figure}

In our truncation, the three-body sector contains a single, pointlike $\phi^{*}\psi^{*}\phi\psi$ term with a coupling $\lambda_{AD}$, which is assumed to vanish in the UV. It is build up by quantum fluctuations during the RG flow and the corresponding Feynman diagrams of the flow equation for $\lambda_{AD}$ is shown in Fig. \ref{FigFlows}(b). First we investigate the unitarity point, $E_\psi=a^{-1}=0$. For this limit we are able to obtain an analytical solution for the flow equation of $\lambda_{AD}$ while away from unitarity we have to rely on a numerical solution.

At the unitarity point all intrinsic length scales drop out of the problem and the system becomes classically scale invariant. At unitarity, the Yukawa coupling $h$ is dimensionless and the only (extrinsic) length scale present is the inverse ultraviolet  cutoff $\Lambda^{-1}$, which defines the validity limit of our effective theory.\\
In our approximation the dimer field $\phi$ develops a large anomalous dimension $\eta=-\frac{\partial_t A_{\phi}}{\bar{A}_{\phi}}=1$ at unitarity which is consistent with the exact solution of the two-body sector \cite{DKS, MFSW}. As the atom and dimer propagators have vanishing gaps in the IR the two-body sector respects a continuous scaling symmetry. 

In order to find the solution of the three-body sector we switch to the rescaled, dimensionless coupling $\tilde\lambda_{AD}\equiv\frac{k^2}{h^2}\lambda_{AD}$. One finds that the flow equation for $\tilde \lambda_{AD}$ becomes independent of $k$ and $h$,
\begin{eqnarray} \label{u1}
\partial_t \tilde\lambda_{AD}&=&\underbrace{\frac{24}{25}\left(1-\frac{\eta}{15} \right)}_{a}\tilde\lambda_{AD}^{2}\underbrace{-\frac{14}{25}\left(1-\frac{4\eta}{35} \right)}_{b}\tilde\lambda_{AD}\notag \\
&+&\underbrace{\frac{26}{25}\left(1-\frac{\eta}{65} \right)}_{c}.
\end{eqnarray}
As was demonstrated in \cite{MFSW,Moroz}, the behavior of the solution of this type of flow equation is determined by the sign of the discriminant $D$ of the right hand side of Eq. (\ref{u1}) which is $D=b^2-4ac<0$. Eq. (\ref{u1}) can be solved analytically and one finds
\beq \label{u2}
\tilde\lambda_{AD}(t)=\frac{-b+\sqrt{-D}\tan\left(\frac{\sqrt{-D}}{2}(t+\delta)\right)}{2a},
\eeq
where $\delta$ is connected to the three-body parameter and determines the initial condition. Most remarkably, the three-body sector exhibits a quantum anomaly: The RG flow of the renormalized coupling $\tilde\lambda_{AD}$ exhibits a limit cycle, which, due to its periodicity, breaks the classically continuous scaling symmetry to the discrete subgroup $Z$. The Efimov parameter can be determined from the period of the limit cycle \cite{MFSW} and is given in our approximation by
\beq \label{u3}
s_{0}=\frac{\sqrt{-D}}{2}\approx0.925203.
\eeq
The exact result is given by $s_0\approx 1.00624$ \cite{BH}. Considering the simplicity of our pointlike approximation, which, as discussed in Sec. \ref{twobody}, does not resolve any momentum or frequency structure of the effective ($k$ dependent) interaction vertex of the three-body sector, the agreement is quite good. In fact, in previous work \cite{MFSW} we have shown how to obtain the exact value of $s_0$ using the FRG.\\

The presence of N-body bound states leads to divergencies in the corresponding N-body vertices. The periodic divergencies in the analytical solution of $\tilde\lambda_{AD}$ in Eq. \eqref{u2} correspond therefore to the presence of the infinitely many Efimov trimer states at the unitarity point.

We can use the latter correspondence to calculate the bound state spectrum also away from unitarity. The trimer binding energies are calculated by determining the atom energies $E_\psi$ for which $\tilde\lambda_{AD}$ exhibits divergencies in the IR as function of $a^{-1}$. The trimer binding energy is then given by $E_T=-3 E_\psi$. The result is shown in Fig. \ref{4BSpectrum}. In this plot the dashed, green lines indicate the binding energies of the Efimov trimer states. For calculational purposes we switch to the static trimer approximation which is completely equivalent to our two-channel model in Eq. \eqref{m2}. We describe this procedure in Appendix \ref{Ap:Trimer}.

At the unitarity point the trimer binding energies form a geometric spectrum and the ratio between adjacent levels is given by
\beq\label{s0eq}
\frac{E_T^{(n+1)}}{E_T^{(n)}}=e^{-\frac{2\pi}{s_0}},
\eeq
which can be understood from the limit cycle flow of $\lambda_{AD}$. At each scale $k=\Lambda e^{t}$, where $\lambda_{AD}$ diverges, one hits a trimer state. The RG scale $k$ can in turn be connected to the atom energy $E_\psi$ \cite{FSMW,MFSW,FSW} and as the divergencies appear periodically in $t$ one easily obtains Eq. \eqref{s0eq}.

There is an additional universal relation obeyed by the trimer energy levels which we may take as a measure of the quality of our approximation. It is given as the relation between the trimer binding energy $E^*$ for $a\to\infty$ and value of $a$ for which the trimer becomes degenerate with the atom-dimer ($a^*_+$) and three-atom threshold ($a^*_-$), respectively. For comparison we define a wave number $\kappa^*$ by $E^*=-\hbar^2\kappa^{*2}/M$ (in our convention, $E^*=-2\kappa^{*2}$) and find
\beq
a^*_-\kappa^*\approx -1.68,\quad a^*_+\kappa^*\approx0.08
\eeq
which has to be compared with the exact result $a^*_-\kappa^*=-1.56(5)$, $a^*_+\kappa^*=0.07076$ from the fully momentum-dependent calculation in \cite{Bedaque,BH}. The agreement with our approximate solution suggests that our model should provide a solid basis for the step to the four-body problem.\\

\section{The four-body sector}\label{4B}
Recently, the solution of the four-body problem in the low-energy limit has gained a lot of interest. In quantum mechanical calculations the existence of two tetramer (four-body bound) states was conjectured for each of the infinitely many Efimov trimers \cite{Platter04,Hammer07}. By calculating the lowest few sets of bound state levels von Stecher et al. \cite{Stecher09, DIncao09} concluded that both ratios of energies between the different tetramers and the trimer state approach universal constants. However, with the quantum mechanical approach the calculation directly at the unitarity point ($a^{-1}=E_\psi=0$), marked explicitly in Fig. \ref{4BSpectrum}, turns out to be difficult, although this point is of great interest when one wants to gather evidence for universality of the four-body system. In fact, in the three-body sector the infinite RG limit cycle appears only exactly at the unitarity point and its universal appearance is directly connected to the breaking of the continuous scale symmetry. Within our approach, the unitarity region is easily accessible.\\
In order to investigate the four-body sector we include all possible, U(1) symmetric, momentum-independent interaction couplings in the effective flowing action $\Gamma_k$. If one assumes all these couplings to be zero at the microscopic UV scale $\Lambda$, one can show, by evaluating all possible Feynman diagrams and using the vacuum hierarchy described in \cite{MFSW}, that from all possible four-body couplings only the three couplings $\lambda_\phi$, $\beta$, and $\gamma$ are built up by quantum fluctuations and are therefore included in Eq. \eqref{m2}. Couplings other than $\lambda_\phi$, $\beta$, and $\gamma$, such as, for instance, the term $\sim (\psi^*\psi)^4$ are not generated during the RG evolution. This consideration leads to our ansatz for the effective average action \eqref{m2}.

For the investigation of the unitarity point we first switch to rescaled, dimensionless couplings

\begin{equation}\label{ren4B}
\tilde\lambda_\phi=\frac{k^3}{\pi^2 h^4}\lambda_\phi,\quad\tilde\beta=\frac{k^4}{h^3}\beta,\quad
\tilde\gamma=\frac{\pi^2 k^5}{h^2}\gamma,
\end{equation}

\begin{center}
\begin{figure*}[t]
\begin{center}
\includegraphics[width=\textwidth]{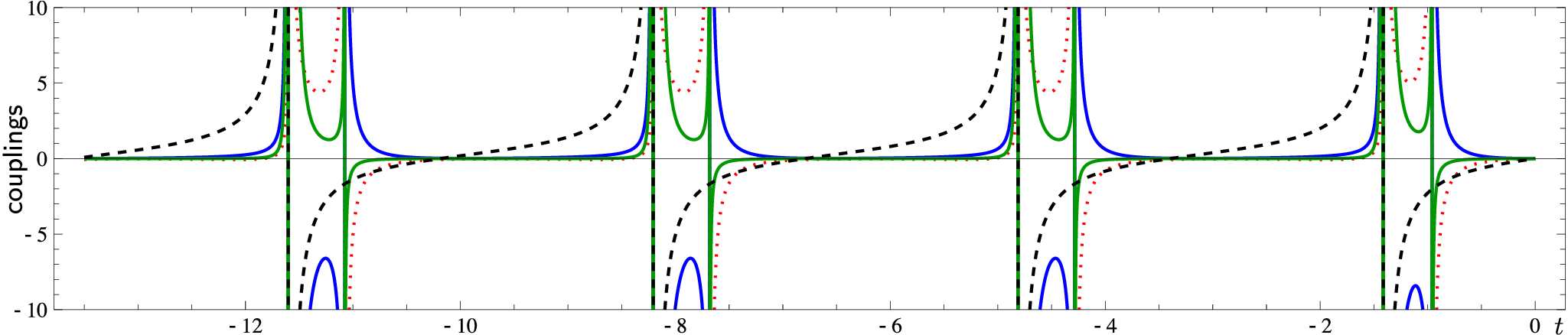}\\
\end{center}
\caption{(Color online) Renormalization group limit cycle behavior of the three- and four-body sector at the unitarity point $E_\psi=a^{-1}=0$. The real parts of the rescaled, dimensionless couplings $\tilde\lambda_{AD,1}$, $4\tilde\lambda_{\phi,1}$, $ \tilde\beta_1/6$, and $\tilde \gamma_1/1000$ are plotted as functions of $t=\ln (k/\Lambda)$. Not only the three-body coupling $\tilde\lambda_{AD,1}$ (dashed, black) exhibits a limit cycle behavior, but also the four-body sector couplings $\tilde\lambda_{\phi,1}$ (red, dotted), $\tilde\beta_1$ (blue, solid), and $\tilde\gamma_1$ (green, solid) obey a limit cycle attached to the three-body sector with the same period.}
\label{4BLC}
\end{figure*}
\end{center}
and obtain the corresponding flow equations by inserting the effective flowing action $\Gamma_k$, Eq. \eqref{m2}, into the Wetterich equation \eqref{m1}. By the use of the rescaled couplings we find three coupled ordinary differential equations, which are again coupled to the two- and three-body sectors, but become explicitly independent of $h$ and $k$. We show the diagrammatic representation of the flow equations in Fig. \ref{FigFlows}(c). Their analytical form at the unitarity point is given by\footnote{For illustrative purpose we show the analytical form of the flow equations at the unitarity point only. Away from this limit their explicit expressions become much more complex.}
\begin{eqnarray}
\partial_t \tilde\lambda_{AD}&=&\frac{128}{125}-\frac{62}{125}\tilde\lambda_{AD}+\frac{112}{125}\tilde\lambda_{AD}^2,\\
\partial_t \tilde\lambda_{\phi}&=&\frac{1}{16}+\frac{1}{3}\tilde\beta-\frac{1}{6}\tilde\lambda_{AD}+3 \tilde\lambda_\phi +\frac{128}{15}\tilde\lambda_\phi^2,\\
\partial_t \tilde\beta&=&\frac{188}{125} \tilde\beta +\frac{1}{6}\tilde\gamma+\frac{128 }{125}\tilde\lambda_{AD}\notag\\
&+&\frac{224}{125}\tilde\lambda_{AD}\tilde\beta-\frac{156}{125}\tilde\lambda_{AD}^2+\frac{4384 }{375}\tilde\lambda_\phi\notag\\
&+&\frac{128 }{15}\tilde\beta \tilde\lambda_\phi -\frac{3968 }{375}\tilde\lambda_{AD}\tilde\lambda_\phi, \\
\partial_t \tilde\gamma&=&\frac{4592}{375}+\frac{8768}{375}\tilde\beta+\frac{128 }{15}\tilde\beta^2\notag\\
&+&\frac{1}{125}\tilde\gamma-\frac{79072}{1875}\tilde\lambda_{AD}-\frac{7936}{375}\tilde\beta\tilde\lambda_{AD}\notag\\
&+&\frac{448}{125}\tilde\gamma\tilde\lambda_{AD}+\frac{74368}{1875}\tilde \lambda_{AD}^2-\frac{5376}{625}\tilde\lambda_{AD}^3.
\end{eqnarray}\\

We pointed out in the last section that the appearance of bound states is connected with divergent vertex functions $\Gamma_k^{(n)}$ and we exploit this behavior to determine the bound state spectrum of the three- and four-boson system. At this point we must note that these infinities are complicated to handle in a numerical solution of the theory. In particular, the numerical treatment of unbounded limit cycles is problematic due to the periodic infinities during the RG flow. In order to circumvent this difficulty we used the method of complex extension, developed in \cite{Moroz}. The basic idea is to extend the domain of the running couplings to the complex plane
\begin{eqnarray} \label{u4}
 \lambda_{AD}\to \lambda_{AD,1}+i\lambda_{AD,2}&\quad&\lambda_{\phi}\to \lambda_{\phi,1}+i\lambda_{\phi,2}\nonumber \\
  \beta\to \beta_{1}+i\beta_{2}&\quad& \gamma\to \gamma_{1}+i\gamma_{2}.
\end{eqnarray}

On the one hand this effectively doubles the number of real flow equations and additional initial conditions must to provided. We choose $\lambda_{AD,2}=\epsilon=10^{-11}$ in our numerical calculation and take all other imaginary parts to be zero in the UV. On the other hand this procedure allows us to perform the numerical integration of the flow equations as it regularizes the periodic infinities in the flow and makes the numerical treatment feasible. Physically, by the complex extension we convert the stable bound states into metastable resonances and by taking different values of $\epsilon$ we are able to vary the decay width of the resonances. One may compare this with the procedure of Braaten and Hammer \cite{Braaten01} who introduce a parameter $\eta^*$ in order to model the decay of the trimers to deeply bound states which have not been included in the effective model. In this line we also view our complex extension as a way to include these deeply bound states in the FRG calculation. Specifically, we find that for $\epsilon\ll 1$ the decay width of the n$^{\textrm{th}}$ Efimov trimer $\Gamma_{\textrm{T}}^{(n)}$ is given by $\Gamma_{\textrm{T}}^{(n)}=4\epsilon E_{\textrm{T}}^{(n)}$ at unitarity. This is in agreement with the result in \cite{BH}
\beq
\Gamma_{\textrm{T}}^{(n)}\approx\frac{4\eta^*}{s_0} E_{\textrm{T}}^{(n)}
\eeq
which holds for small $\eta^*$. Thus, for $\epsilon \ll1$, the relation to the parameter $\eta^*$ introduced by Braaten and Hammer is given by
\beq
\epsilon=\frac{\eta^*}{s_0}.
\eeq
\begin{figure}[t!]
\begin{center}
\includegraphics[width=\columnwidth]{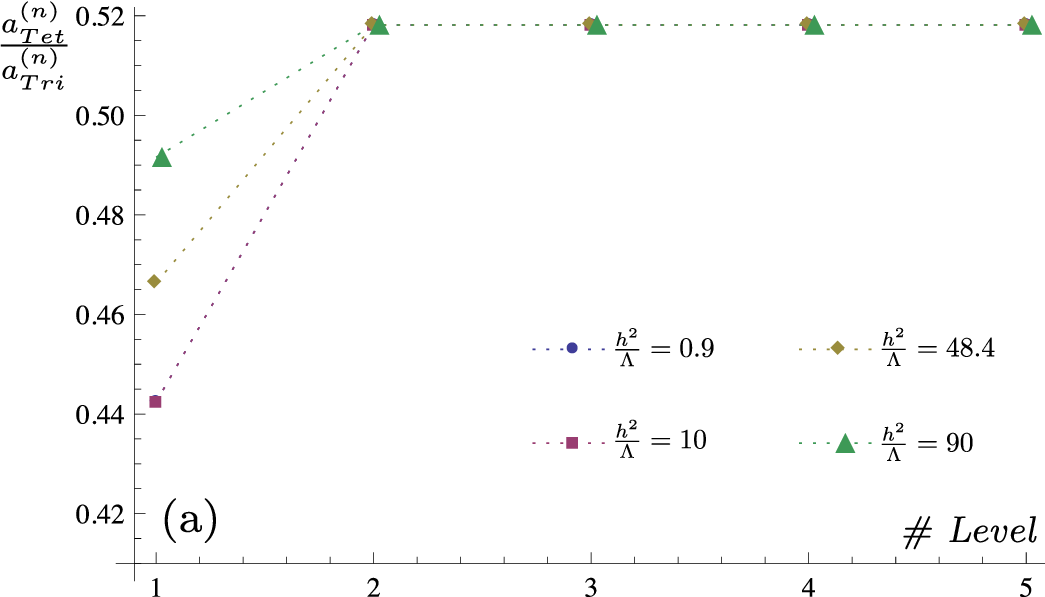}\\
\vspace{6mm}
\includegraphics[width=\columnwidth]{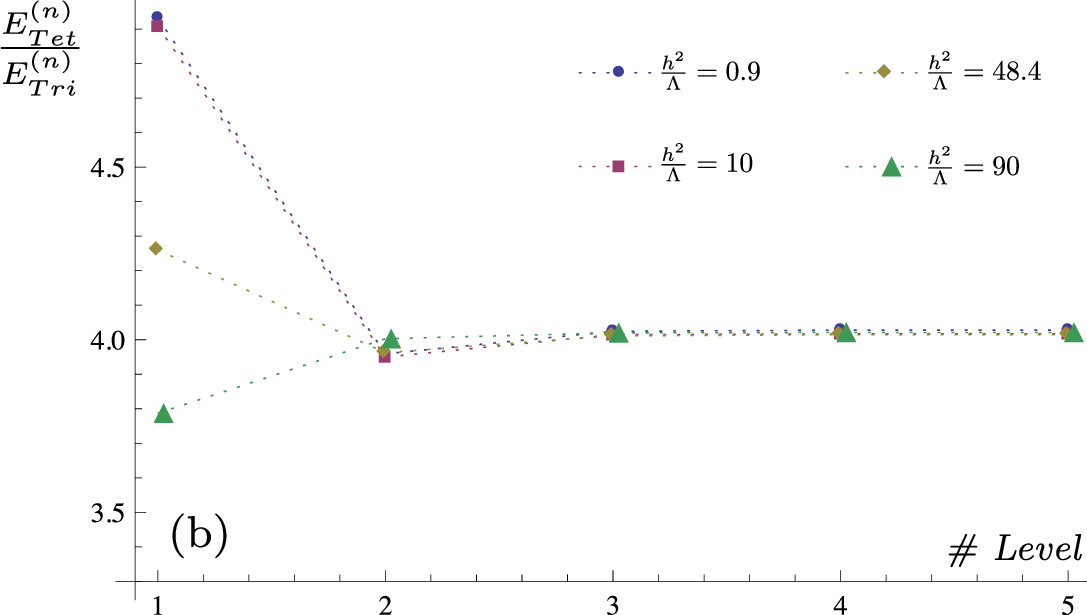}
\end{center}
\caption{(Color online) Calculation of the universal ratios for the lowest five set of levels. The calculation is done for different values of the Yukawa coupling $h^2/\Lambda$ which determines the effective range in our model. The dotted lines are only guide for the eye. (a) Ratios between the values of scattering lengths $a_{Tet}^{(n)}$ and $a_{Tri}^{(n)}$ for which the tetramer and corresponding trimer become degenerate with the four-atom threshold. (b) Ratios between the values of binding energies $E_{Tet}^{(n)}$ and $E_{Tri}^{(n)}$ at resonance $a\to\infty$. }
\label{4BRat}
\end{figure}
The result of the numerical calculation of the four-body sector at unitarity is shown in Fig. \ref{4BLC}. Here, we display the RG flows of the real parts of all nonzero three- and four-body sector couplings as a function of the RG scale $t=\ln (k/\Lambda)$. The three-body coupling $\tilde\lambda_{AD,1}$ (black dotted line) exhibits the well-known limit cycle behavior, described in Section \ref{STBS}, with the period being connected to the Efimov parameter $s_0$. Remarkably, there is an additional limit cycle in the flow of the four-body sector couplings with a periodic structure of exactly the same frequency as the three-body sector. This four-body sector limit cycle exhibits resonances which are shifted with respect to the ones of the three-body system. The magnitude of this shift is given by a new universal number, which is inherent to the four-body sector.\\

Our observation is that the four-body sector is intimately connected with the three-body sector at the unitarity point. It is permanently attached to the running of the three-body sector from the first three-body resonance on. From here on the periodic structure of the flow remains unchanged as one goes to smaller values of $k$. Due to this tight bond between the three- and four-body sector, there stays no room for an additional four-body parameter.\\
We also find that the magnitude of the shift beyond the first resonance is neither dependent on the initial values of the four-body sector couplings in the UV nor is it influenced by finite range corrections which we are able to check by choosing different values for the Yukawa coupling $h$. Arbitrary choices lead to the same behavior. Having done this calculation directly at the unitarity point our conclusion is, that, within our simple approximation, the four-body sector behaves truly universal and independent of any four-body parameter confirming the conjecture made by Platter et al. and von Stecher et al.. We expect that universality will also hold for an improved truncation.\\

Naively one expects that each resonance in the flow of the vertex functions is connected to the presence of a bound state. As one observes there are also additional resonances in the four-body sector being degenerate with the three-body sector resonances. However, we arrive at the conclusion that these resonances are artifacts of our approximation. The mathematical structure of the flow equations is of a kind that divergencies in the three-body sector directly lead to a divergent four-body sector. We are confident that the resonances at these positions will disappear as one includes further momentum dependencies in the field theoretical model. Therefore we can already infer from the calculation at unitarity that within our approximation we are only able to resolve a single tetramer state attached to each trimer state also away from unitarity. In contrast, the ``exact'' quantum mechanical calculations in \cite{Platter04,Hammer07,Stecher09,DIncao09} predict the existence of two tetramer states which have recently been observed by Ferlaino et al. \cite{Ferlaino09}. As one includes further momentum dependencies, it is well possible that not only the degenerate resonance disappears but also new, genuine resonances associated with the ``missing'' tetramer state will appear at the same time. This effect indeed occurs in the three-body problem. There, it is essential to include the momentum dependent two-atom vertex. Only under this condition one arrives at the quadratic equation as in \eqref{u1}  which gives rise to the Efimov effect. This can easily be seen by taking a look at the flow equation of $\lambda_{AD}$ depicted as Feynman diagrams in Fig. \ref{FigFlows}(b). The assumption of a momentum independent two-atom interaction corresponds to a momentum (and frequency) independent dimer propagator. In this approximation the first term on the RHS of Fig. \ref{FigFlows}(b) vanishes because all poles of the loop frequency integration lie on the same complex frequency half-plane. This directly leads to the loss of the Efimov effect in this crude level of approximation.\\
We can also use our model to investigate the full bound state energy spectrum by solving the flow equations for arbitrary values of the scattering length $a$. The energy levels of the various bound states are then determined by varying the energy of the fundamental atoms $E_\psi$ such that one finds a resonant four-body coupling in the IR. The result of this calculation, using the static trimer approximation presented in Appendix \ref{Ap:Trimer}, is shown in Fig. \ref{4BSpectrum}, where we plot the energy levels of the various bound states versus the inverse scattering length. We find one tetramer state attached to each of the Efimov trimer states. These tetramer states become degenerate with the four-atom threshold for negative scattering length and merge into the dimer-dimer threshold for positive $a$. In the experiment this leads to the measured resonance peaks in the four-body loss coefficient. In order not to overload the plot we show only the first three sets of levels, although the FRG method allows to calculate an arbitrary number of them. One also observes that the shape of the tetramer levels follows the shape of the trimer levels. In analogy to the three-body sector one can calculate a universal formula relating a tetramer binding energy $E^*_{Tet}=-2\kappa_{T}^{*2}$ at $a\to\infty$ with the corresponding scattering length at which the tetramer becomes degenerate with the four-atom threshold $a^*_{T-}$ and the dimer-dimer threshold $a^*_{T+}$, respectively. We find
\beq
a^*_{T-}\kappa^*_{T}\approx-1.75,\quad a^*_{T+}\kappa^*_{T}\approx0.20.
\eeq
In their recent quantum mechanical calculations von Stecher et al. were able to calculate the lowest few sets of bound state energy levels \cite{Stecher09}. From their behavior it was inferred that the ratio between the tetramer and trimer binding energies approaches a universal number within these first few sets of levels. Figuratively speaking it is therefore expected that the universal regime in the energy plot in Fig. \ref{4BSpectrum}  is reached very fast as one goes to smaller $a^{-1}$ and $E_\psi$. 

In order to investigate this observation we calculate the behavior of two ratios as a function of the set of level for which they are determined. The first ratio relates the negative scattering lengths $a^{(n)}_{Tet}$ and $a^{(n)}_{Tri}$ for which the n$^\textrm{th}$ tetramer and trimer become degenerate with the four-atom threshold. The second is the ratio between the binding energies of the n$^\textrm{th}$ tetramer $E^{(n)}_{Tet}$ and the n$^\textrm{th}$ trimer $E^{(n)}_{Tri}$ at resonance, $a\to\infty$. The resulting plots are shown in Fig. \ref{4BRat}. We calculate the ratios for different values of the microscopic couplings in order to test the degree of universality of the various sets of energy levels. In the plots we show in particular the dependence on the choice of the Yukawa coupling $h$ determining the effective range $r_{\textrm{eff}}$ of the model. As one sees, only the first of the ratios depend on the microscopic details. Already from the second set of levels on the microscopic details are washed out and the ratios become independent of the choice of initial conditions: The regime of universality is reached extremely fast and as $a^{-1}$ and $E_\psi$ are lowered one will ultimately find the four-body limit cycle described above.\\
For the asymptotic ratios we find
\begin{eqnarray}
a^{(n)}_{Tet}&\approx&0.518\,a^{(n)}_{Tri},\\
E^{(n)}_{Tet}&\approx&4.017\,E^{(n)}_{Tri}.
\end{eqnarray}
Von Stecher et al. find $a^{(n)}_{Tet}/a^{(n)}_{Tri}\approx0.43\,(0.9)$ for the deeper (shallower) bound tetramer and $E^{(n)}_{Tet}/E^{(n)}_{Tri}\approx4.58\,(1.01)$, respectively. Considering the simplicity of our model the agreement is quite good.\\
With an ultracold bosonic $\textrm{Cs}$ gas Ferlaino et al. found $a^{(n)}_{Tet}/a^{(n)}_{Tri}\approx0.47\,(0.84)$. In this remarkable experiment only the lowest set of tetramer states in the energy spectrum had been accessible due to the particular scattering length profile. Considering our observation that the deepest set of levels is still strongly dependent on the microscopical details it cannot be expected to find the universal numbers in this particular setting. Therefore more experiments for bosons interacting via a larger scattering lengths would be desirable.\\

\section{Conclusions}\label{Conclusion}
In this paper, we investigated the four-body problem with the help of the functional renormalization group. Employing a simple two-channel model with pointlike three- and four-body interactions we were able to investigate universal properties at the unitarity point $a\to\infty$, $E_\psi=0$ as well as to perform computations away from it.\\
In the RG language the Efimov physics of the three-body problem manifests itself as an infinite RG limit cycle behavior of the three-body coupling constant at unitarity. We found that also the four-body sector is governed by such a limit cycle which is solely induced by the RG running of the three-body sector, signaling the absence of a four-body parameter.\\
We also computed the energy spectrum away from unitarity and were able to obtain the universal relations between four- and three-body observables in our approximation. Our calculation provides an explanation for the findings of von Stecher et al. \cite{Stecher09}, who found that these ratios approach universal constants very quickly as they are computed for higher and higher excited states. We also found a dependence of the ratios for the lowest level on microscopic details such as the effective range. This in turn is of relevance for the experimental observations by Ferlaino et al. \cite{Ferlaino09}. In this experiment the lowest states have been measured and one can therefore not expect to find the exact universal relations between them.\\
Considering the simplicity of our model, the agreement with the previous studies in \cite{Platter04,Hammer07,Stecher09} is quite good. There had been some disagreement in literature about universality and the absence or existence of a four-body parameter, see e.g. \cite{Yamashita06,Adhikari95,Amado73}. Our RG results support the conclusion that the four-body system is universal and independent of any four-body parameter.\\
An important shortcoming of the pointlike approximation is the absence of the shallower of the two tetramer states. Obviously the pointlike approximation of the three- and four-body sectors is not sufficient and in future work one should include momentum dependent interactions. From the energy spectrum in Fig. \ref{4BSpectrum} it becomes also evident that the excited tetramer states can decay into an energetically lower lying trimer plus atom. The higher excited states in the four-body system are therefore expected to have an intrinsic finite decay width \cite{Hammer07}. Whether this width has a universal character still remains an opened question as well as in which way the corresponding imaginary coupling constants will change the RG analysis.\\
The inclusion of the full momentum dependencies in the three- and four-body sector seems to be a rather complicated task. In the effective field theory study of the three-boson system the introduction of a dynamical dimer field, often called the di-atom trick \cite{BH}, has been a decisive step towards the exact solution of the three-body problem. From this perspective we suggest that the inclusion of a dynamical trimer field in the effective action might help to simplify the momentum dependent calculation.\\
The four boson system remains still a subject with many open questions. With our RG analysis in the pointlike approximation, we made the first step towards a renormalization group description of the four-body problem supplementing the previous quantum mechanical approaches. From this perspective this work provides a starting point for a deeper understanding of universality in the four-body problem.

\section{Acknowledgments}
We thank E. Braaten, S. Floerchinger, C. H. Greene, H. W. Hammer, J. Pawlowski, J. von Stecher, C. Wetterich, and W. Zwerger for stimulating discussions. We are indebted to F. Ferlaino, R. Grimm, and S. Knoop for many insightful discussions and for pointing out the problem to us in the first place. RS thanks the DFG for support within the FOR 801 `Strong correlations in multiflavor ultracold quantum gases'. SM is grateful to KTF for support.

\appendix

\section{The trimer approximation and rebosonization}\label{Ap:Trimer}

In this appendix we will apply the rebosonization method developed in \cite{Bosonization} to our model \eqref{m2}. We introduce an additional trimer field $\chi$, representing the bound state of three bosons, which then mediates the atom-dimer interaction. A similar procedure had already been used long time ago by Fonseca and Shanley \cite{Fonseca76} in the context of nuclear physics and was recently employed by us in \cite{FSMW,FSW} for the treatment of the three-component Fermi gas. There are several reasons for employing this procedure. First, it is useful to reduce the number of resonances one has to integrate through in the RG flow. Instead of calculating the divergent coupling $\lambda_{AD}$ one only has to calculate zero-crossings of the trimer energy gap which is numerically much easier to handle. Secondly, by the introduction of a dynamical trimer field one may be able to mimic some of the complicated momentum structure of the atom-dimer interaction in a simple way which could probably be sufficient to find the missing tetramer state in our calculation. The third point is of a more technical nature and concerns the method of rebosonization, which we will employ here in a quite extensive manner.

In the three-body sector the real atom-dimer coupling $\lambda_{AD}$ exhibits divergencies when the energy gap of the fundamental atoms $E_\psi$ is tuned such that one hits the trimer bound state in the IR. In the static trimer approximation, the coupling $\lambda_{AD}$ is mediated by the exchange of a trimer field $\chi$ with the non-dynamical, inverse propagator $P_\chi=m_\chi^2$, which can be depicted as 
\begin{center}
\includegraphics[width=0.8\columnwidth]{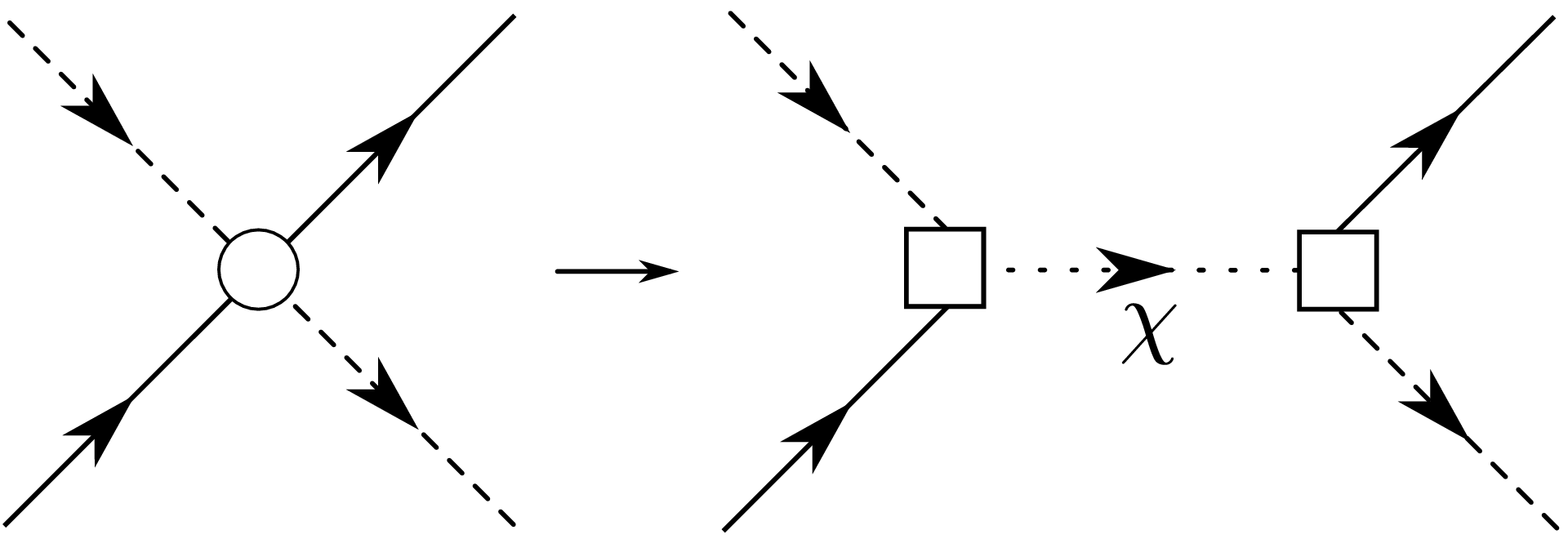}
\end{center}
The trimer field $\chi\sim\psi^3$ is introduced on the microscopic scale by a Hubbard-Stratonovich transformation and our ansatz for the effective average action, motivated by the resulting classical action, reads
\begin{eqnarray}\label{ThreeChannel}
\Gamma_k&=&\int_{x}\{\psi^{*}(\partial_{\tau}-\Delta+E_\psi)\psi\notag\\
&&+\phi^{*}\left(A_{\phi}(\partial_{\tau}-\frac{\Delta}{2})+m_{\phi}^{2} \right)\phi+\chi^{*}m_{\chi}^{2} \chi\notag\\
&&+\frac{h}{2}(\phi^{*}\psi\psi+\phi \psi^{*}\psi^{*})+\lambda_{AD}\phi^{*}\psi^{*}\phi\psi \notag\\
&&+g (\chi^*\phi\psi+\chi\phi^*\psi^*)\notag\\
&&+\lambda_{\phi}(\phi^{*}\phi)^{2}+\beta(\phi^{*}\phi^{*}\phi\psi\psi+\phi\phi\phi^{*}\psi^{*}\psi^{*})\notag\\
&&+ \gamma \phi^{*}\psi^{*}\psi^{*}\phi\psi\psi +\delta_1 \chi^*\psi^*\chi\psi\notag\\
&&+\delta_2(\chi^*\psi^*\phi\phi+\chi\psi\phi^*\phi^*)\notag\\
&&+\delta_3(\chi^*\psi^*\phi\psi\psi+\chi\psi\phi^*\psi^*\psi^*)\}.
\end{eqnarray}
The Yukawa interaction $g$ couples the trimer field to the dimer and atom field. The $\delta_i$ are the additional U(1) symmetric four-body couplings which are generated by quantum fluctuations. All other possible couplings can be shown to stay zero during the RG evolution provided they are zero at the UV scale. Also the coupling $\lambda_{AD}$ is regenerated through a box diagram in the RG flow. However, it is possible to absorb all these emerging couplings by the use of the rebosonization procedure.\\
For this matter we promote the trimer field $\chi$ to be explicitly scale dependent, $\chi\to\chi_k,\,\chi^*\to\chi^*_k$ and the Wetterich equation generalizes to
\begin{eqnarray}
\nonumber
\partial_k \Gamma_k[\Phi_k]&=&\frac{1}{2}\text{Tr}\, \left(\Gamma_k^{(2)}[\Phi_k]+R_k\right)^{-1}\partial_k R_k\\ 
&& + \left(\frac{\delta}{\delta \Phi_k}\Gamma_k[\Phi_k]\right)\partial_k \Phi_k,
\label{eq:genwetteq}
\end{eqnarray}
where $\Phi_k$ now includes all fields including the trimer fields $(\chi,\chi^*)$. The additional term in the generalized flow equation \eqref{eq:genwetteq} allows for the absorption of the reemerging couplings since one has the freedom to choose the scale dependence of the trimer fields as a function of fields. In order to continuously eliminate the couplings $\lambda_{AD}$ and $\delta_i$ we choose
\begin{eqnarray}\label{eq:kdepchi}
\partial_k\chi_k&=&\phi\psi\zeta_{a,k}+\psi^*\chi_k\psi\zeta_{b,k}\notag\\
&&+\psi^*\phi\phi\zeta_{c,k}+\psi^*\phi\psi\psi\zeta_{d,k},\notag\\
\partial_k\chi^*_k&=&\phi^*\psi^*\zeta_{a,k}+\psi\chi_k^*\psi^*\zeta_{b,k}\notag\\
&&+\psi\phi^*\phi^*\zeta_{c,k}+\psi\phi^*\psi^*\psi^*\zeta_{d,k}.
\end{eqnarray}

Upon inserting Eq. \eqref{eq:kdepchi} into the generalized Wetterich equation \eqref{eq:genwetteq} the condition that the flows of  $\lambda_{AD}$ and $\delta_i$ vanish leads to
\begin{eqnarray}\label{ap1}
\zeta_a&=&-\frac{\partial_k \lambda_{AD}}{2 g},\quad\zeta_b=-\frac{\partial_k \delta_{1}}{2 m_\chi^2}\notag\\
\zeta_c&=&-\frac{\partial_k \delta_{2}}{ m_\chi^2},\quad\zeta_d=-\frac{\partial_k \delta_{3}+g\zeta_b}{m_\chi^2}.
\end{eqnarray}
When one calculates now the flow equations of the remaining flowing couplings by projecting Eq. \eqref{eq:genwetteq} onto them, one obtains new contributions due to the presence of additional terms arising from Eq. \eqref{eq:kdepchi}.\\ 
In our static trimer approximation the trimer field has no dynamical propagator and the model given by Eq. \eqref{ThreeChannel} is completely equivalent to the two-channel model in Eq. \eqref{m2}. Furthermore, no regulator has to be specified for the trimer field since in our approximation the original atom-dimer coupling $\lambda_{AD}$ is solely replaced by $g^2/m_\chi^2$. At this point it already becomes clear why the modified flow equations will be easier to handle numerically: instead of calculating a divergent $\lambda_{AD}$ in the three-body sector one has only to deal with zero crossings of $m_\chi^2$ at the values of $k$ where originally $\lambda_{AD}$ had divergencies. The modified flow equations are given by
\begin{eqnarray}
\partial_t g&=& \partial_t g|_{\Phi_k}+m_\chi^2 \zeta_a,\notag\\
\partial_t \beta&=& \partial_t \beta |_{\Phi_k}+g \zeta_c,\notag\\
\partial_t \gamma&=& \partial_t \gamma |_{\Phi_k}+2 g \zeta_d,
\end{eqnarray}  
where the first terms in the flows are the original flow equations with the trimer field taken to be scale independent. In fact, by expressing all flow equations in terms of the coupling $G\equiv g^2$ one can also get rid off the problematic $g$ in the denominator of $\zeta_a$ in Eq. \eqref{ap1}. We point out that the static trimer approximation allows to calculate easily the three-body sector. In the four-body sector the original divergencies of $\lambda_{AD}$ still appear since trimers $\chi$ appear in the corresponding flow diagrams and therefore one has to deal with terms $\sim 1/m_\chi^2$. For this reason it had been essential to perform the complex extension described in Section \ref{4B}. Finally, the bound state spectrum of the trimers can be computed by calculating the poles of the trimer propagator, $m_\chi^2(k=0,E_\psi)=0$, in a straightforward manner.

%%%%%%%%%%%%%%%%%%%%%%%%%%%%%

\end{document}